\documentclass[preprint,tightenlines,floatfix,aps,a4paper]{revtex4-1}

\usepackage{graphicx}
\usepackage{amssymb}
\usepackage{amsmath}
\usepackage{color}
\usepackage{psfrag}
\usepackage{epsfig}
\usepackage{dsfont}

\newcommand{\ket}[1]{| #1 \rangle}

\newcommand{\rb}[1]{\left( #1 \right)}

\newcommand{\ew}[1]{\langle #1 \rangle}
\newcommand{\beq}{\begin{eqnarray}}
\newcommand{\eeq}{\end{eqnarray}}

\newcommand{\op}[2]{| #1 \rangle \langle #2 |}

\newcommand{\eq}[1]{Eq.~(\ref{#1})}
\newcommand{\fig}[1]{Fig.~\ref{#1}}

\newcommand{\bs}[1]{\boldsymbol{#1}}

\begin{document}
\title{Delayed feedback control in quantum transport}
\author{Clive Emary}
\affiliation{
  Institut f\"ur Theoretische Physik,
  Hardenbergstr. 36,
  TU Berlin,
  D-10623 Berlin,
  Germany
}

\date{\today}
\begin{abstract}
  Feedback control in quantum transport has been predicted to give rise to several interesting effects, amongst them quantum state stabilisation and the realisation of a mesoscopic Maxwell's daemon.  These results were derived under the assumption that control operations on the system be affected instantaneously after the measurement of electronic jumps through it.
  In this contribution I describe how to include a delay between detection and control operation in the master equation theory of feedback-controlled quantum transport. 
  I investigate the consequences of delay for the state-stabilisation and Maxwell's-daemon schemes.  Furthermore, I describe how delay can be used as a tool to probe coherent oscillations of electrons within a transport system and how this formalism can be used to model finite detector bandwidth.
\end{abstract}

\maketitle
%%%%%%%%%%%%%%%%%%%%%%%%%%%%%%%%%%%%%%%%%%%%%%%%%%%%%%%%%%%%%%%%%%%%%%%%%

The feedback control of quantum transport has recently been predicted to give rise to several interesting effects such as the freezing of current fluctuations \citep{Brandes2010}, stabilisation of quantum states \citep{Kiesslich2011,Poeltl2011} and the realisation of a mesoscopic Maxwell's daemon \citep{Schaller2011}.
In all these schemes, the feedback was of Wiseman-Milburn type, in which the control operations are performed instantaneously and directly after quantum jumps of the system \citep{Wiseman1994,Wiseman2009}.
The aim of this contribution is to analyse the effects of delay in the feedback control of quantum transport.  In particular, we are interested in when the control operations follow the jump processes not directly, but rather after some time delay, be it originating from the finite-response time of feedback hardware, or introduced deliberately into the control loop.

Away from a transport setting, delay in quantum feedback has been considered by a number of authors, e.g. \citet{Nishio2009,Combes2010,Combes2011,Amini2012}.  In particular, \citet{Wiseman1994} considered, at a formal level, delay in Wiseman-Milburn control and has derived a delayed-feedback quantum master equation (QME) in the limit of small delay time. 
Here I rederive this delayed QME using an alternative approach that makes it clear that this equation can actually be valid for arbitrary delay, provided one assumes that if a jump occurs within the delay time of a previous control operation, then the interrupted control operation is skipped.  
This delayed control scheme results in a non-Markovian master equation which, by construction, has well-behaved solutions.
To enable the calculation of transport properties, I generalise this result to make connection with full-counting statistics (FCS) of electron transfer \citep{Levitov1996} by inclusion of counting fields (see \citet{Poeltl2011} for overview of this procedure without delay).

With this formalism in place, I investigate the consequences of feedback delay for the state-stabilisation of \citet{Poeltl2011} and Maxwell's daemon of \citet{Schaller2011}.  
As an example of how delay in control need not necessarily be a negative thing \citep{Schoell2008,Schoell2009},  I describe how a deliberately-swept delay can be used as a tool to probe coherent oscillations of electrons within a transport system . 
Finally, I show briefly how this delay formalism can be used to model a finite-bandwidth electron-counting detector.

%%%%%%%%%%%%%%%%%%%%%%%%%%%%%%%%%%%%%%%%%%%%%%%%%%%%%%%%%%%%%%%%%%%%%%%%%
%%%%%%%%%%%%%%%%%%%%%%%%%%       FORMALISM        %%%%%%%%%%%%%%%%%%%%%%%
%%%%%%%%%%%%%%%%%%%%%%%%%%%%%%%%%%%%%%%%%%%%%%%%%%%%%%%%%%%%%%%%%%%%%%%%%
\section{Delayed feedback in the quantum master equation \label{SEC:FORM}}

Let us consider a transport system described by the QME, $\dot \rho = \mathcal{W}\rho$, with $\rho = \rho(t)$ the reduced density matrix of the system (e.g. quantum dot) at time $t$ and superoperator $\mathcal{W}$ the Liouvillian of the system  \citep{Brandes2005}.  Let the Liouvillian be decomposed as $\mathcal{W}=\mathcal{W}_0 + \mathcal{J}$, where $\mathcal{J} = \sum_\alpha \mathcal{J}_\alpha$ describes quantum jump processes --- in particular those in which electrons enter and leave the system ---, and $\mathcal{W}_0$ describes the evolution without jumps.  In terms of the Hilbert space jump operators $L_\alpha$, we have 
$
  \mathcal{W}_0\rho 
  = 
  -i\left[H,\rho\right] 
  -   \frac{1}{2}
  \left\{
    \sum_\alpha L^\dag_\alpha L_\alpha
    ,\rho
  \right\}
$
and  
$
  \mathcal{J}_\alpha\rho  = L_\alpha \rho L_\alpha^\dag
$, such that $\mathcal{W}$ is of Lindblad form.

\begin{figure}[thb]
  %\begin{center}
    \includegraphics[width=0.99\columnwidth,clip=true]{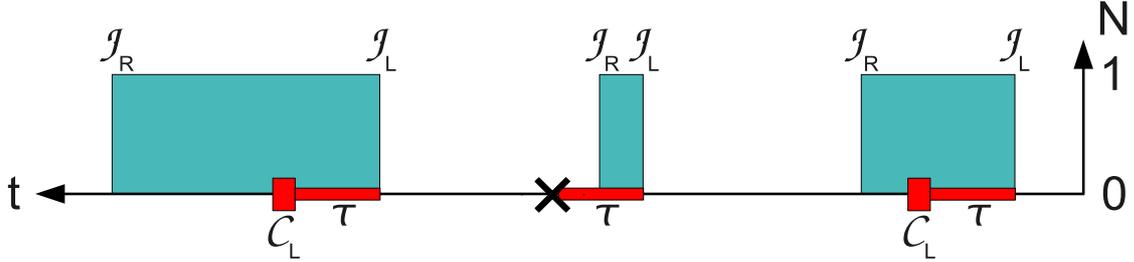}
    \caption{
      A sketch of a portion of a quantum trajectory of a transport system coupled to two leads. Time flows from right to left (the same direction as the operators act in \eq{CON:traj}), jump processes from the left lead ($\mathcal{J}_L$) transfer electrons to the system, whereas as jumps to the right lead ($\mathcal{J}_R$) remove electrons.  
      In the strong Coulomb blockade regime, the system occupation $N$ correspondingly switches between 0 and 1.
      With delayed feedback in effect, the control operations $\mathcal{C}_L$ are applied a time $\tau$ after every left jump except when that jump is followed by the next within the delay time.  In this latter case, the control operation is skipped.  This ``control-skipping'' is depicted for the middle pair of jumps, where the cross indicates that no control operation is applied.
    \label{FIG:jumps}
    }
  %\end{center}
\end{figure}

An intuitive picture of feedback control can be obtained by considering the solution of the QME in terms of quantum trajectories \citep{Carmichael1993}:
\beq
  \rho(t) = \sum_{n=0}^\infty\int^{t}_{0}dt_n \ldots \int_{0}^{t_{2}}dt_{1} 
  \underbrace{
    \Omega_0(t_{n}-t_{n-1}) 
    \mathcal{J}
    \ldots
    \mathcal{J}
    \Omega_0(t_{2}-t_{1}) 
    \mathcal{J}
    \Omega_0(t_1) 
  }_{n~\mathrm{jumps}}
  \rho(0)
  \label{CON:traj}
  ,
\eeq
with $\Omega_0(t) = e^{\mathcal{W}_0t}$.
A sketch of a trajectory for transport through a quantum dot is shown in \fig{FIG:jumps}.

%%%%%%%%%%%%%%%%%%%%%%%%%%%%%%%%%%%%%%%%%%%%%%%%%%%%%%%%%%%%%%%%%%%%%%%%%
%%%%%%%%%%%%%%%%%%%%%%%%%%%%%%%%%%%%%%%%%%%%%%%%%%%%%%%%%%%%%%%%%%%%%%%%%

Feedback control can be added to this scheme by considering that after each jump $\mathcal{J}_\alpha$ we operate on the system with control operator $\mathcal{C}_\alpha$, assumed instantaneous.  In standard Wiseman-Milburn control, each jump is followed immediately with the control operation. The trajectories including control operations can then be resummed, such that the density matrix evolves under the action of the modified control Liouvillian $\mathcal{W}_\mathrm{C}=\mathcal{W}_0 + \sum_\alpha \mathcal{C}_\alpha \mathcal{J}_\alpha$.  

We now want to consider a delay between jump $ \mathcal{J}_\alpha$ and corresponding control operation $ \mathcal{C}_\alpha$.  It would seem a simple matter to simply insert the control operations into each trajectory a time $\tau$ after each jump. However, in doing so, one encounters a problem: what happens when a second jump occurs within the delay time of the first?  
If we assume that the control operation is applied regardless of whether  this occurs or not, the trajectories become complicated and can not be resummed as a master equation. 
To avoid this, we assume that in the case where a jump occurs in the delay time of a previous jump, the control operation of the first jump is simply skipped.  In other words, if the time between two subsequent jumps is shorter than the delay time, no control operation for the first jump is performed.
This I shall call the {\em control-skipping} assumption.
This assumption is a plausible: it is not hard to imagine that the feedback electronics be reset by the arrival of a second jump as they prepare to enact the control operation of the previous. Certainly, in the case where the delay is introduced deliberately, this feedback scheme could be arranged.  

With this assumption, addition of delayed feedback control is effected by the replacement of the no-jump propagators following jumps $\mathcal{J}_\alpha$ as
\beq
  \Omega_0(t)\mathcal{J}_\alpha \to
  \left\{
  \begin{array}{cl}
    \Omega_0(t)\mathcal{J}_\alpha & t < \tau_\alpha\\
     \Omega_0(t-\tau_\alpha)\mathcal{C}_\alpha\Omega_0(\tau_\alpha)\mathcal{J}_\alpha 
       & t \ge \tau_\alpha\\
  \end{array}
  \right.
  ,
\eeq
throughout all trajectories.  In the first instance, time between the jumps is too short for the delayed control to be implemented, in the second line, there is sufficient time and the control operation is implemented.  For the sake of generality, we have included here a subscript on the delay time such that each operation may have its own associated delay.

With this replacement, the controlled trajectories can still be resummed such that the density matrix in Laplace space reads
\beq
  \rho(z) = 
  %\sum_n [\Omega_0(z)  \sum_\alpha\mathcal{D}_\alpha(z) \mathcal{J}_\alpha]^n\Omega_0(z)\rho_0
  \int_0^\infty dt e^{-z t } \rho(t)
  = \frac{1}{z - \mathcal{W}_0 - \sum_\alpha \mathcal{D}_\alpha(z) \mathcal{J}_\alpha}\rho_0
  ,
\eeq
with delayed-control superoperator
\beq 
  \mathcal{D}_\alpha(z) = \mathds{1} + (\mathcal{C}_\alpha-\mathds{1})e^{(\mathcal{W}_0-z)\tau_\alpha}
  \label{Dz}
  .
\eeq
Translating back into time-domain, we obtain the nonMarkovian QME
\beq
  \dot\rho(t) = \int_0^t dt' 
  \left[\frac{}{}
    \mathcal{W}_0\delta(t-t') +  \sum_\alpha\mathcal{D}_\alpha(t-t') \mathcal{J}_\alpha
  \right]
  \rho(t')  
  ,
\eeq 
with $
   \mathcal{D}_\alpha(t) = \delta(t) + (\mathcal{C}_\alpha-1)e^{\mathcal{W}_0\tau_\alpha}\delta(t-\tau_\alpha)
$.
Evaluating the delta-functions we obtain the delayed QME:
\beq
  \dot\rho(t) =
    \mathcal{W}\rho(t) 
    + \sum_\alpha
      (\mathcal{C}_\alpha-1) e^{\mathcal{W}_0\tau_\alpha}\mathcal{J}_\alpha
      \theta(t-\tau_\alpha)\rho(t-\tau_\alpha) 
  \label{NMMEtime}
    ,
\eeq
in which the time evolution of the density matrix $\rho(t)$ depends not only on the state of the system at time $t$ but also at previous times $\left\{t-\tau_\alpha\right\}$.
The $\theta$-functions that accompany the delayed terms mean that, in accordance with the construction of this master equation from its specific solution, the history of the system up to time $t=0$ is not needed. 
Furthermore, knowledge of the explicit solution allows us to conclude that no positivity or normalisation issues arise with this particular nonMarkovian QME.
\eq{NMMEtime} is a slight generalisation of the form given in \citet{Wiseman1994}.  There, this expression was derived in the framework of the stochastic Schr\"odinger equation as being valid only to first order in the delay time(s).
The foregoing shows that \eq{NMMEtime} is actually valid for arbitrary delay, provided the additional control-skipping assumption is made.

To facilitate the calculation of the FCS of transport processes, we introduce the counting fields $\left\{\chi_\alpha\right\}$ associated with tunneling of electrons into/out of leads $\left\{\alpha\right\}$. With counting fields, and assuming that transport into each lead is unidirectional (infinite bias limit), the nonMarkovian Laplace-space delayed-control Liouvillian reads
\beq
   \mathcal{W}_{DC}(\chi,z) 
   &=& 
  \mathcal{W}_0 
  + \sum_\alpha
    \mathcal{D}_\alpha(\chi,z) 
    \mathcal{J}_\alpha e^{i\chi_\alpha}
    \label{WDCtwoJ}
    ,
\eeq
with delayed control operation
\beq
  \mathcal{D}_\alpha(\chi,z) 
  &=&  
  \mathds{1} + \left[\mathcal{C}_\alpha(\chi)-\mathds{1}\right]e^{(\mathcal{W}_0-z)\tau} 
  \label{DztwoJ}
  ,
\eeq
where $\chi$ without subscript refers to the complete set of counting fields.
Note that in general the control operations $\mathcal{C}_\alpha$, and hence $\mathcal{D}_\alpha$, can transfer electrons and thus depend on the counting fields.  
Calculating the transport properties of nonMarkovian QMEs is well understood, see e.g. \citet{Flindt2008, Emary2011}.  Providing that we start counting at $t=0$, no inhomogeneous term is required in the QME.

%%%%%%%%%%%%%%%%%%%%%%%%%%%%%%%%%%%%%%%%%%%%%%%%%%%%%%%%%%%%%%%%%%%%%%%%%
%%%%%%%%%%%%%%%%%%%%%%%%       STABILISATION      %%%%%%%%%%%%%%%%%%%%%%%
%%%%%%%%%%%%%%%%%%%%%%%%%%%%%%%%%%%%%%%%%%%%%%%%%%%%%%%%%%%%%%%%%%%%%%%%%
\section{Nonequillibrium state stabilisation \label{SEC:STAB}}

As first application, let us re-analyse the feedback stabilization protocol of \citet{Poeltl2011}, this time with delay.  The system consists of a double quantum dot (DQD) described by the three states: `empty' $\ket{0}$ and left- and right- occupied states, $\ket{L}$ and $\ket{R}$.  The Hamiltonian of the DQD reads 
$
  H_\mathrm{DQD}  = \frac{1}{2}\epsilon \sigma_z +T_C\sigma_x 
$
with pseudospin operators $\sigma_z \mathord=\op{L}{L}\mathord-\op{R}{R}$, $\sigma_x\mathord= \op{L}{R}\mathord+\op{R}{L}$. The two transport processes are tunneling into the DQD from the left lead,  $L_L = \sqrt{\Gamma_L} \op{L}{0}$, and out to the right, $L_R = \sqrt{\Gamma_R} \op{0}{R}$. We consider only the $\epsilon=0$ case in the following.

In \citet{Poeltl2011}, the control operation was chosen as a coherent qubit rotation in $x$-$z$ plane conditioned on the tunnel of an electron into the DQD:
\beq
  \mathcal{C}_L = 
  \exp
  \left\{
    \theta_C 
    [\sin \theta \Sigma_x 
    +
      \cos \theta \Sigma_z
      ]
  \right\}
  ;\quad 
  \mathcal{C}_R = 0,
\eeq
with rotations induced by Pauli matrices $\Sigma_\alpha \rho = -i\left[\sigma_\alpha,\rho\right]$.
By choosing control parameters $\theta_C$ and $\theta$ correctly,
this control operation was used to rotate the state of the incoming electron into an eigenstate of the effective Hamiltonian 
$
\widetilde{H} = H_\mathrm{DQD} - \frac{i}{2}
    \sum_\alpha L^\dag_\alpha L_\alpha
$, a state protected from further evolution until the next jump.  In the limit $\Gamma_L\to \infty$ where state $\ket{0}$ can be eliminated (the ``transport qubit'' limit), the system effectively spends all its time in this state, which is thus stabilised.

%%%%%%%%%%%%%%%%%%%%%%%%%%%%%%%%%%%%%%%%%%%%%%%%%%%%%%%%%%%%%%%%%%%%%%%%%
\begin{figure}[tb]
  \begin{center}
    \includegraphics[width=0.65\columnwidth,clip=true]{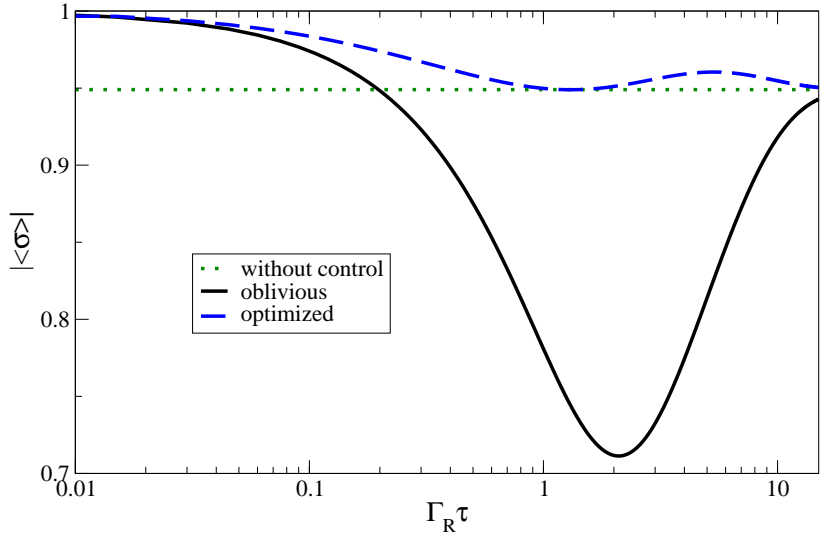}
    \caption{
    The length of the stationary Bloch-vector of the DQD transport qubit with feedback control as a function of delay time $\tau$.
    Results for two strategies of choosing control parameters are shown here: oblivious (black solid line) and optimised (blue dashed line).  The value of $|\ew{\bs \sigma}|$ without feedback control is shown (green dotted line).
    The parameters were $\epsilon=0$ and $T_c = 0.24\Gamma_R$.
    \label{FIG:stab1}
    }
  \end{center}
\end{figure}
%%%%%%%%%%%%%%%%%%%%%%%%%%%%%%%%%%%%%%%%%%%%%%%%%%%%%%%%%%%%%%%%%%%%%%%%%

Introducing delay through the application of \eq{WDCtwoJ} and \eq{DztwoJ} we obtain the delayed-control kernel
\beq
   \mathcal{W}_{DC}(\chi,z) 
   &=& 
  \mathcal{W}_0 
  +\mathcal{J}_R e^{i\chi_R}
  + \mathcal{D}_L(z) 
    \mathcal{J}_L e^{i\chi_L}
    ,
\eeq
with a delayed control operation that does not depend on counting fields
\beq
  \mathcal{D}_L(z) 
  &=&  
  \mathds{1} + (\mathcal{C}_L-\mathds{1})e^{(\mathcal{W}_0-z)\tau}
  .
\eeq
The central question is to what degree can the stationary state of the system (obtained from $\mathcal{W}_{DC}(\chi=0,z=0)\rho_\mathrm{stat}=0 $) be purified by feedback in the presence of delay.
To this end we consider three strategies for choosing the control parameters:
\begin{itemize}
  \item We simply use the values from the stabilisation protocol without delay, a strategy dubbed `oblivious control' in \citet{Combes2011};
  \item We choose the angles such that the purity is maximised for a given delay time;
  \item Without delay,  state stabilisation is correlated with Poissonian statistics of the transport process.  We can therefore attempt to use this same the criterion to set the control parameters in the presence of delay.
\end{itemize}

Results for the first two strategies are shown in \fig{FIG:stab1}, where we plot as a function of delay time the length of the Bloch vector of the stationary state, a measure of the states purity ($|\sigma|=1$ for a pure state).
Delay clearly serves to reduce the purity of the end state. With oblivious control at small delay times, the length of the Bloch vector drops linearly:
\beq
  |\ew{\bs\sigma}| \sim
  \left\{
  \begin{array}{cc}
    1 - 
    \rb{
      \Gamma_R - 8T_c^2/\Gamma_R - \kappa
    }\tau
      &  4T_C < \Gamma_R\\
    1 - \Gamma_R \tau/2
      &  4T_C > \Gamma_R\\
  \end{array}
  \right.
  ,
\eeq
where the two cases arise from a bifurcation in the nature of the stabilisable states without delay.
In the presence of delay, feedback-control only offers a purity improvement over the non-controlled steady-state for delays $\Gamma_R\tau \lesssim 0.2$.  The reduction of the purity is strongest around $\tau\Gamma_R \sim 1$.  For large delays $\tau\Gamma_R \gg 1$, the stationary state with control reverts to that without since, in this limit, the probability that successive jumps occur within the delay time is high and the control operation is only very rarely enacted.  

Explicitly choosing the control parameters to maximise the purity shows a marked increase in the purity over the oblivious approach. As \fig{FIG:stab1} shows, the the purity of the controlled state with strategy is significantly lies above that of its uncontrolled counterpart for most values of the delay.

%%%%%%%%%%%%%%%%%%%%%%%%%%%%%%%%%%%%%%%%%%%%%%%%%%%%%%%%%%%%%%%%%%%%%%%%%
\begin{figure}[tb]
  \begin{center}
    \includegraphics[width=1\columnwidth,clip=true]{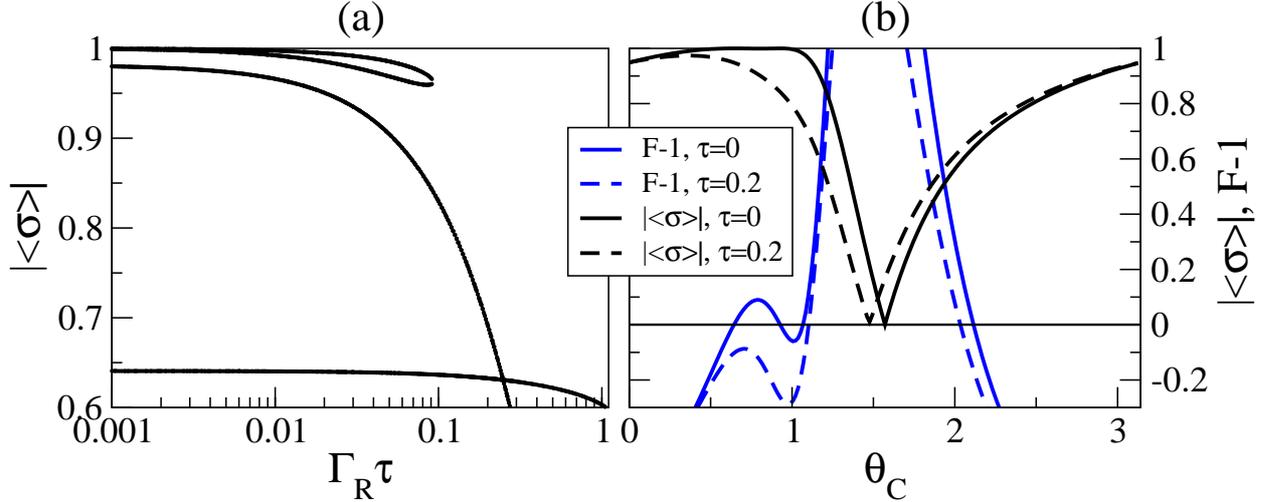}
    \caption{
      State stabilisation with control parameters based on the FCS.
      {\bf (a)}: Length of the Bloch vector as a function of delay time.  Here the control angle $\theta$ was fixed at at a value $\theta=\pi/2$ (the non-delayed choice) and angle $\theta_C$ was chosen such that the Fano factor of transferred charge was exactly unity.  For each value of the delay, there are several choices of control angles which give $F=1$ and hence multiple solutions are plotted. The topmost two solutions correspond to the two stabilised pure states for $\tau\to 0$.  Only for $\Gamma_R\tau\lesssim 0.1$ do there exist control parameters for which $F-1$ is zero and the stationary state has a high purity.
      {\bf (b)}:  This behaviour can be explained by considering the length of Bloch vector and Fano factor as functions of control angle $\theta_C$ (with  $\theta=\pi/2$).  Results are plotted for zero and finite ($\tau=0.2\Gamma_R$) delays.  At zero delay, the function $F-1$ crosses the zero axis in four places; two of these solutions correspond to maximum purity states; the remaining two, low purity states. For larger delays, $\tau\gtrsim 0.1\Gamma_R$, the leftmost maximum of $F-1$ drops below zero and the two high-purity unit-Fano-factor solutions disappear.
      Parameters were as \fig{FIG:stab1}
    \label{FIG:stab2}
    }
  \end{center}
\end{figure}
%%%%%%%%%%%%%%%%%%%%%%%%%%%%%%%%%%%%%%%%%%%%%%%%%%%%%%%%%%%%%%%%%%%%%%%%%

\fig{FIG:stab2} investigates the strategy of choosing the control angles based on the  the FCS.   Here, we just look for a (shotnoise) Fano factor $F$ equal to unity (the Poissonian value) to adjudge this.
For each value of $\tau$ there exists multiple choices of control parameters that give $F=1$.  For small delay times $\Gamma\tau \ll 1$, two of these solutions have high purity and are continuous with the stabilisable states of the $\tau=0$ system.  For small delays, relying on the FCS to locate useful control parameters remains a valid strategy. 
However, above a certain delay time, $\tau \gtrsim 0.1$ in \fig{FIG:stab2}a, these high-purity solutions disappear and selecting for $F=1$ drives the system into one of two highly mixed states. This behaviour is explained in \fig{FIG:stab2}b.  Away from the small $\tau$ limit then, the zeroes of $F-1$ are unrelated with any kind of purity optimisation and this criterion should be avoided.

%%%%%%%%%%%%%%%%%%%%%%%%%%%%%%%%%%%%%%%%%%%%%%%%%%%%%%%%%%%%%%%%%%%%%%%%%
%%%%%%%%%%%%%%%%%%%%%%%%         M DAEMON         %%%%%%%%%%%%%%%%%%%%%%%
%%%%%%%%%%%%%%%%%%%%%%%%%%%%%%%%%%%%%%%%%%%%%%%%%%%%%%%%%%%%%%%%%%%%%%%%%
\section{Maxwell's Daemon\label{SEC:MD}}

\citet{Schaller2011} described how a single-electron transistor (SET) with feedback can act as a Maxwell's Daemon and transfer charge against a voltage gradient whilst no net work is performed on the system.  Several feedback schemes were discussed in this work, but here I shall just discuss `Scheme IIa' (details below) since this is of the appropriate type for our delay treatment.

The SET model consists of a quantum dot with just two states: `empty', $\ket{0}$, and `full', $\ket{1}$, connected to two leads at finite bias and temperature.  A quantum point contact is used to monitor the charge state of the dot and without feedback, the system Liouvillian reads
\beq
  \mathcal{W}(\chi_L,\chi_R)
  =
  \mathcal{W}_0 + \mathcal{J}_I (\chi_L,\chi_R) + \mathcal{J}_O(\chi_L,\chi_R)
\eeq
where counting fields $\chi_{L,R}$ keep track of electron movements through the left and right barriers of the dot.  
Since the QPC can only tell us the occupation of the dot (but not e.g. from which lead the electron has tunneled), the two jump super-operators on which the control scheme is based are $\mathcal{J}_I $ and $\mathcal{J}_O$, which describe inward and outward jumps respectively.
In a the basis $\left\{\ket{0},\ket{1}\right\}$ we have
\beq
  \mathcal{W}_0
  =
  \sum  \Gamma_\alpha F_\alpha^0
  ;~~
  \mathcal{J}_I (\chi_L,\chi_R)
  = \sum_\alpha \Gamma_\alpha F_\alpha^- e^{-i \chi_\alpha}
  ;~~
  \mathcal{J}_O(\chi_L,\chi_R)
  = \sum_\alpha \Gamma_\alpha F_\alpha^+ e^{i \chi_\alpha}
  ;\!\!\!
  \nonumber\\
  F_\alpha^0 = 
  \rb{
  \begin{array}{cc}
    -f_\alpha & 0 \\
    0 & - (1-f_\alpha)
  \end{array}
  }
  ;~
  F_\alpha^+ = 
  \rb{
  \begin{array}{cc}
    0 & (1-f_\alpha) \\
    0 & 0
  \end{array}
  }
  ;~
  F_\alpha^- = 
  \rb{
  \begin{array}{cc}
    0 & 0 \\
    f_\alpha & 0
  \end{array}
  }
  ,~~
\eeq
where $\Gamma_\alpha$ is the tunnel rate associated with lead  $\alpha=L,R$, and where 
$f_\alpha$ is the corresponding Fermi function, $f_\alpha = [e^{\beta(\epsilon-\mu_\alpha)}+1]^{-1}$ with $\epsilon$ the dot level energy,$\mu_\alpha$ the chemical potential of lead $\alpha$ and $\beta=1/kT$ the inverse thermal energy. 

The control scheme IIa of \citet{Schaller2011} is, on detection of a tunnel event either into or out of the system, to change the barrier heights and return them instantaneously.  The corresponding control operations read
\beq
  \mathcal{C}_{I/O} (\chi_L,\chi_R) 
  &=& 
  \exp\rb{
    \sum_\alpha
    \delta_\alpha^{I/O}
    \rb{
      F_\alpha^0 + F_\alpha^- e^{-i \chi_\alpha} + F_\alpha^+ e^{i \chi_\alpha}
    }
  },
\eeq
where parameters $\delta_\alpha^{I/O}$ describe the `strength' of the feedback transition involving lead $\alpha$ given an in/out jump.  For simplicity, let's use the ``maximum feedback'' case and set $\delta_R^I=\delta_L^O=\delta\to \infty$ and $\delta_L^I=\delta_R^O=0$.  We will also consider a symmetric SET, $\Gamma_L=\Gamma_R=\Gamma$.
With delay, then, the controlled Liouvillian reads
\beq
  \mathcal{W}_{DC}(\chi_L,\chi_R,z)
  =
  %\mathcal{W}_0 + \mathcal{D}_I (\chi_L,\chi_R,z) \mathcal{J}_I (\chi_L,\chi_R) 
  %+ \mathcal{D}_O (\chi_L,\chi_R,z) \mathcal{J}_O(\chi_L,\chi_R)
  \mathcal{W}_0 
  + \sum_{\alpha=I/O}
    \mathcal{D}_\alpha (\chi_L,\chi_R,z) \mathcal{J}_\alpha (\chi_L,\chi_R)
\eeq
with
\beq
  \mathcal{D}_{I/O} (\chi_L,\chi_R,z) 
  &=& 
  \mathds{1} +  
  \rb{\mathcal{C}_{I/O} (\chi_L,\chi_R)- \mathds{1} } 
  e^{
  (\mathcal{W}_0 - z)\tau
  }
  .
\eeq
These equations are slightly different to \eq{WDCtwoJ}and \eq{DztwoJ} since here we have finite bias and bidirectional tunneling. The inclusion of delay is directly analogous, however.

%%%%%%%%%%%%%%%%%%%%%%%%%%%%%%%%%%%%%%%%%%%%%%%%%%%%%%%%%%%%%%%%%%%%%%%%%
\begin{figure}[tb]
  \begin{center}
    \includegraphics[width=\columnwidth,clip=true]{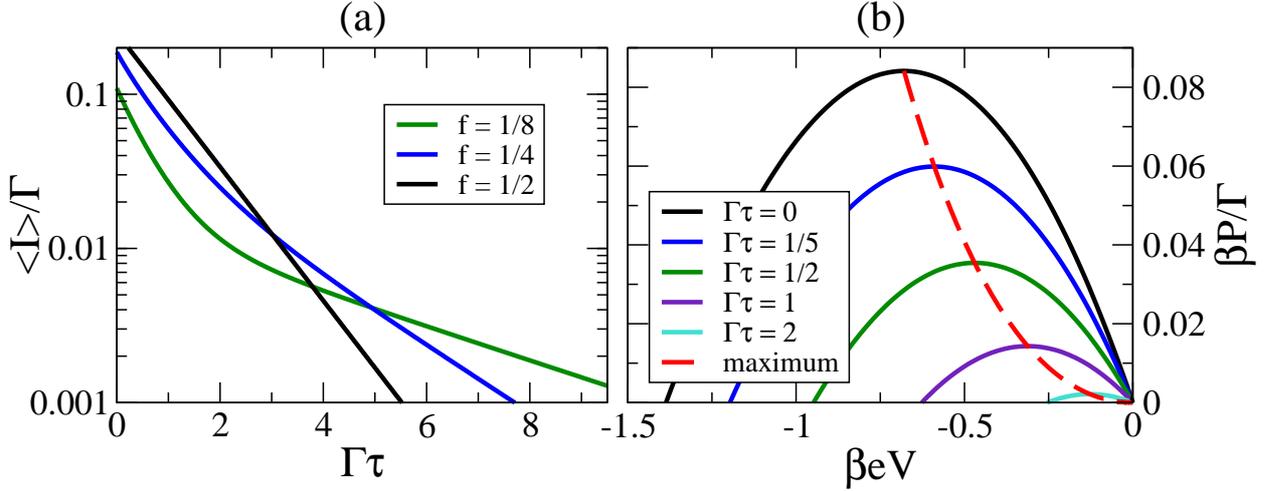}
    \caption{
      {\bf(a)}
      Equillibrium current through the Maxwell daemon SET with delayed feedback as a function of delay time $\tau$.  Results for several values of the Fermi function $f$ are shown.  For $f=1/2$, the decay of the current with $\tau$ is purely exponential.
      {\bf(b)}
      Power generated by the Maxwell daemon SET as a function of bias $eV$ applied symmetrically about the dot level: $\mu_L-\epsilon=\epsilon-\mu_R=eV/2$.  Results for several values of the delay $\tau$ are shown (solid lines) and the maximum power point indicated (red dashed line).
      In both cases, the parameters were $\delta_R^I=\delta_L^O=\delta \to \infty$, $\delta_L^I=\delta_R^O=0$ and $\Gamma_L=\Gamma_R=\Gamma$. 
    \label{FIG:MD}
    }
  \end{center}
\end{figure}
%%%%%%%%%%%%%%%%%%%%%%%%%%%%%%%%%%%%%%%%%%%%%%%%%%%%%%%%%%%%%%%%%%%%%%%%%

Even with delayed feedback, the current through the SET can be obtained analytically. 
Let us first consider equillibrium conditions such that $f_L=f_R =f$.  In this case, the current through the SET is
\beq
  \ew{I} = 
  \Gamma f(1-f)
  \frac{
    fe^{2 \Gamma \tau} - 2 f(1-f) e^{2 f \Gamma \tau} + (1-f) e^{4f \Gamma \tau }
  }{
    e^{2(1+f)\Gamma \tau}-f(1-f)( e^{2\Gamma \tau} + e^{4f\Gamma \tau})
  }.
\eeq
This result is plotted in \fig{FIG:MD}a.  With the dot level placed on resonance with the chemical potential of the leads, we have $f=1/2$ and the current assumes the simple form
\beq
  \ew{I} = \frac{\Gamma}{4}e^{-\Gamma \tau}
  ,
\eeq
such that it is clear that the current is exponentially suppressed by the delay.  For $ \Gamma \tau \ll 1$, however, the induced current remains close to the $\tau=0$ value.
The behaviour for $f\ne 1/2$ is a little more complicated but the basic trend is the same.

Away from equillibrium, the operation of the daemon may be assessed by considering the power generated by the device, $P\equiv - \ew{I} V$.  This power is shown for a symmetric bias configuration in \fig{FIG:MD}b for the case of maximum feedback.  Irrespective of the value of $\tau$, this function shows a single maximum as a function of bias.  Without delay, the maximum power generated by the device is obtained numerically as $P\approx 0.084 \Gamma kT$.  With increasing delay, the maximum power decreases approximately exponentially.  The bias at which this maximum is reached also moves towards zero.  For a delay time $\Gamma\tau = 1$ (which represents a large delay), the maximum power is $P\approx 0.014 \Gamma kT$.

%%%%%%%%%%%%%%%%%%%%%%%%%%%%%%%%%%%%%%%%%%%%%%%%%%%%%%%%%%%%%%%%%%%%%%%%%
%%%%%%%%%%%%%%%%%%%%%%%%         DQD PROBE        %%%%%%%%%%%%%%%%%%%%%%%
%%%%%%%%%%%%%%%%%%%%%%%%%%%%%%%%%%%%%%%%%%%%%%%%%%%%%%%%%%%%%%%%%%%%%%%%%
\section{Probing coherent oscillations with delayed feedback \label{SEC:DQDosc}}

Whilst the previous two examples illustrate the negative effects of delay on previously established feedback schemes, delay may also be used constructively. In this section we investigate coherent oscillations of a DQD with delayed feedback as our probe.  
The model without control is the same as in Sec.~\ref{SEC:STAB}.   The feedback scheme we use is to detect jumps through the left barrier and conditionally apply the control operation $e^{\mathcal{K}_R(\chi_R)}$ with
$
  \mathcal{K}_R(\chi_R)\rho = A
  \rb{
   \op{0}{R}\rho\op{R}{0} e^{i\chi_R}
   - \frac{1}{2}\op{R}{R}\rho
   - \frac{1}{2}\rho \op{R}{R}
  }
$, which constitutes an instantaneous lowering and restoration of the {\em right} barrier.  Parameter $A$ is a measure of the feedback strength. 
%%%%%%%%%%%%%%%%%%%%%%%%%%%%%%%%%%%%%%%%%%%%%%%%%%%%%%%%%%%%%%%%%%%%%%%%%
\begin{figure}[tb]
  \begin{center}
    \includegraphics[width=0.7\columnwidth,clip=true]{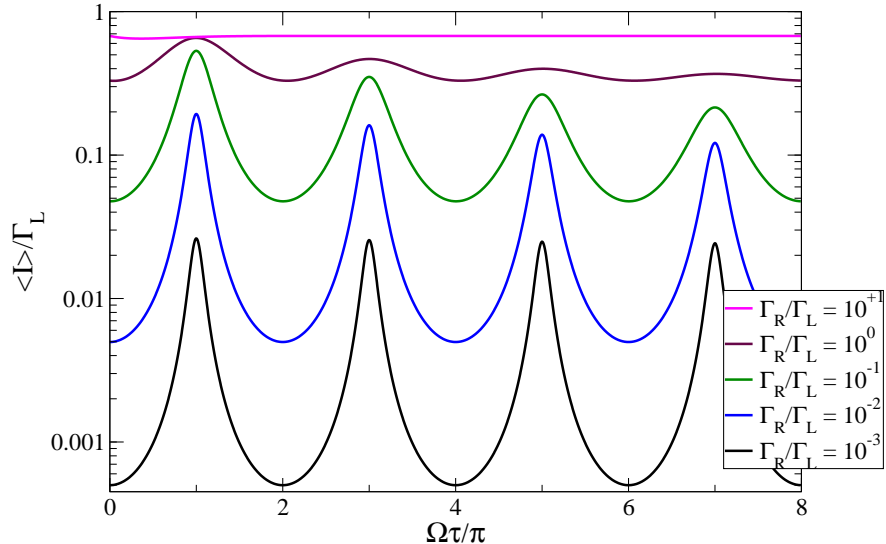}
    \caption{
      Stationary current through feedback-controlled DQD  as a function of delay time, $\tau$. The delayed control scheme is such that a time $\tau$ after an electron tunnel into the system from the left, the right barrier is dropped and returned instantaneously.  A series of peaks in the current is observed that occur when the delay time 
      is equal to odd-integer multiples of $\pi/\Omega$, the time taken for the electron to be coherently transfered across the DQD.
      Parameters: $\epsilon=0$, $T_C=3\Gamma_L$ and dimensionless feedback strength $A=4$
    \label{FIG:OSC}
    }
  \end{center}
\end{figure}
%%%%%%%%%%%%%%%%%%%%%%%%%%%%%%%%%%%%%%%%%%%%%%%%%%%%%%%%%%%%%%%%%%%%%%%%%

\fig{FIG:OSC} shows the stationary current through the DQD as a function of delay time.  When the ratio of right to left tunnel rates is small enough, the current shows a pronounced series of peaks that occur when the delay time is equal to odd-integer multiples of $\pi/\Omega/$. These peaks arise when the control operation is enacted just as the electron has completed a half-integer number of coherent oscillations starting from the left dot-state.
The current consists of two components: that which occurs without control, and an extra component induced by the feedback operation.  As $\Gamma_R$ is decreased, the non-feedback component is reduced and the oscillations become more pronounced. Furthermore, since the only source of dephasing in this model is the coupling to the right lead, decreasing $\Gamma_R$ also decreases this dephasing and this accounts for the decreased damping that accompanies the increased visibility of the oscillations.

It is clear then that this delayed-feedback scheme allows us to image the coherent oscillations taking place within the DQD.  More generally, such schemes provide a way to investigate oscillatory behaviour in other transport systems. In this sense, delayed-feedback plus current measurement can provide an additional method for studying transport dynamics, complementary to the finite-frequency current correlations \citep{Emary2007d,Marcos2010a} or pulsed operation \citep{Hayashi2003}.

%%%%%%%%%%%%%%%%%%%%%%%%%%%%%%%%%%%%%%%%%%%%%%%%%%%%%%%%%%%%%%%%%%%%%%%%%
%%%%%%%%%%%%%%%%%%%%%%%%     FINITE BANDWIDTH     %%%%%%%%%%%%%%%%%%%%%%%
%%%%%%%%%%%%%%%%%%%%%%%%%%%%%%%%%%%%%%%%%%%%%%%%%%%%%%%%%%%%%%%%%%%%%%%%%
\section{Modelling a finite-bandwidth detector\label{SEC:BW}}

The above delay formalism can be adapted to model the effects of a finite-bandwidth detector.  In a generic (unidirectional) Liouvillian without control each jump operator $\mathcal{J}_\alpha$ for which electrons are being counted is immediately followed by a counting-field factor $e^{i\chi_\alpha}$.  This can be interpreted as the detector reacting instantaneously to the occurrence of a system jump.  More realistic is that the detector takes a finite time $\tau$ to react such that jumps go undetected if several occur within this detector reaction time.
This situation is very similar to the delayed-feedback situation described above but instead of having a control operator act at a time $\tau$ after the jump, we have a detection event described by a counting-field factor.  The Liouvillian for this situation may thus be written
\beq
   \mathcal{W}_{BW}(\chi,z) 
   &=& 
  \mathcal{W}_0 
  + \sum_\alpha
    \mathcal{D}_\alpha(\chi_\alpha,z) 
    \mathcal{J}_\alpha 
    \label{WBWtwoJ}
    ,
\eeq
with delayed counting factor
\beq
  \mathcal{D}_\alpha(\chi_\alpha,z) 
  &=&  
  \mathds{1} + (e^{i\chi_\alpha}-1)e^{(\mathcal{W}_0-z)\tau_\alpha} 
  \label{DBWztwoJ}
  .
\eeq
The equations are the same as \eq{WDCtwoJ} and \eq{DztwoJ}, but here the counting-field factors occurs not in \eq{WBWtwoJ} but rather in place on the control operation in \eq{DBWztwoJ}.

As example, let us consider the SET of section \ref{SEC:MD} (without control) in the infinite bias limit $f_L\to 1$, $f_R\to0$.
For symmetric rates, $\Gamma_L=\Gamma_R=\Gamma$, the current detected by counting electrons with a detector reaction time of $\tau$ flowing through the SET reads
\beq
  \ew{I}_\mathrm{detected} = \frac{1}{2}\Gamma_R e^{-\Gamma_R \tau}.
\eeq
This shows an exponential suppression due to the detector lag over the actual current flowing, which is $\ew{I} = \Gamma_R/2$.
With a reliable detector, we should have $\Gamma_R \tau\ll1$ and the detected current reads $\ew{I}_\mathrm{detected} \approx \frac{1}{2}\Gamma_R \rb{1-\Gamma_R \tau}$. 

These results can be compared with the ``detector-state model'' \citep{Naaman2006,Gustavsson2007,Flindt2007}.  Using an additional detector degree of freedom with detector switching rate $\Gamma_D$, the detected current of the symmetric SET was found to be 
\beq
  \ew{I}_\mathrm{detected} =\frac{1}{2} \Gamma_R \frac{k}{1+k}
  ;\quad
  k = \frac{\Gamma_D}{2\Gamma_R}
  \label{avIBF}
\eeq
For a fast detector, $\Gamma_D\gg\Gamma_R$, we may approximate $\ew{I}_\mathrm{detected} \approx \frac{1}{2}\Gamma_R \rb{1-\frac{2\Gamma_R}{\Gamma_D}}$.  Thus, identifying the parameters of these two detector models as $\tau =2/\Gamma_D$, the descriptions of the  behaviour in the experimentally important regime are consistent.  For larger values of $\tau$, the two models differ: the delayed-counting model predicts an exponential decay of the current whereas \eq{avIBF} predicts an algebraic one.

%%%%%%%%%%%%%%%%%%%%%%%%%%%%%%%%%%%%%%%%%%%%%%%%%%%%%%%%%%%%%%%%%%%%%%%%%
%%%%%%%%%%%%%%%%%%%%%%%%    CONCLUSIONS     %%%%%%%%%%%%%%%%%%%%%%%
%%%%%%%%%%%%%%%%%%%%%%%%%%%%%%%%%%%%%%%%%%%%%%%%%%%%%%%%%%%%%%%%%%%%%%%%%
\section{Conclusions \label{SEC:CONC}}

The delay formalism described here contains two effects:  the obvious one that the control operations follow a time $\tau$ after a jump, but also that control operations are rejected when the time between jumps is shorter than the delay time.  This second means that in the limit $\tau\Gamma\to\infty$ (with $\Gamma$ the typical rate of the jump processes) the feedback control is completely frozen out.  Whilst it is interesting to try to relax this second control-skipping assumption, without it, a master equation description (even a nonMarkovian one) is not possible.  Such a control scheme could however be readily simulated.
Extension to piecewise-constant control schemes, such as Scheme I of \citet{Schaller2011} and \citet{Schaller2012} should also be possible.

Delay has been seen here to have a negative impact on the stabilisation and Maxwell daemon schemes, reducing the purity of the stabilised state and the power production, respectively. For small delay times, however, good results are still obtainable.
Interestingly, the quantum stabilisation scheme appears to be impacted more severely than the effectively-classical Maxwell's daemon.  The purification effect based on FCS detection disappears completely for $\Gamma \tau \gtrsim 0.1$, whereas the Maxwell Daemon is capable of producing some power even with a very poor detector $\Gamma\tau >1$. The power drops off exponentially with increasing $\Gamma \tau$, though.

In a more positive sense, we have shown how a deliberately-delayed  control scheme introduces a extra time-scale into the system, which can be used as a probe of the transport dynamics.

Whilst we have concentrated on feedback control of quantum transport, these results should also be applicable to other systems, e.g. quantum optics.

\subsection*{Acknowledgements}
%\ack{
I am grateful to T. Brandes, Gerold Kie\ss{}lich, Christina P\"oltl, and Gernot Schaller for useful discussions. This work was funded by the Deutsche Forschungsgesellschaft through SFB-910
%}

%%%%%%%%%%%%%%%%%%%%%%%%%%%%%%%%%%%%%%%%%%%%%%%%%%%%%%%%%%%%%%%%%%%%%%%%%
%%%%%%%%%%%%%%%%%%%%%      REFERENCES     %%%%%%%%%%%%%%%%%%%%%%%%%%%%%%%
%%%%%%%%%%%%%%%%%%%%%%%%%%%%%%%%%%%%%%%%%%%%%%%%%%%%%%%%%%%%%%%%%%%%%%%%%
%\bibliographystyle{plainnat}

%%%%%%%%%%%%%%%%%%%%%%%%%%%%%%%%%%%%%%%%%%%%%%%%%%%%%%%%%%%%%%%%%%%%%%%%%

\end{document}